\documentclass[prl,twocolumn,showpacs,preprintnumbers,amsmath,amssymb,citeautoscript]{revtex4-1}

\usepackage{graphicx}
\usepackage{dcolumn}
\usepackage{bm}
\usepackage{color}

\begin{document}

\title{Evolution of quasiparticle excitations with critical mass enhancement in superconducting $A$Fe$_2$As$_2$ ($A=$ K, Rb, and Cs)}

\author{Y.~Mizukami$^{1,2}$}
\author{Y.~Kawamoto$^2$}
\author{Y.~Shimoyama$^{2}$}
\author{S.~Kurata$^{1,2}$}
\author{H.~Ikeda$^{2,3}$}
\author{T.~Wolf$^4$}
\author{D.\,A.~Zocco$^4$}
\author{K.~Grube$^4$}
\author{H.~v.\,L\"ohneysen$^4$}
\author{Y.~Matsuda$^2$}
\author{T.~Shibauchi$^{1,2}$}

\affiliation{
$^1$Department of Advanced Materials Science, University of Tokyo, Kashiwa, Chiba 277-8561, Japan\\
$^2$Department of Physics, Kyoto University, Sakyo-ku, Kyoto 606-8502, Japan\\
$^3$Department of Physical Sciences, Ritsumeikan University, Kusatsu, Shiga 525-8577, Japan\\
$^4$Institute of Solid State Physics (IFP), Karlsruhe Institute of Technology, D-76021 Karlsruhe, Germany
}

\date{\today}

\begin{abstract}

In the heavily hole-doped iron-based superconductors $A$Fe$_2$As$_2$ ($A=$ K, Rb, and Cs), the electron effective mass increases rapidly with alkali-ion radius. 
To study how the mass enhancement affects the superconducting state, we measure the London penetration depth $\lambda(T)$ in clean crystals of $A$Fe$_2$As$_2$ down to low temperature $T\sim0.1$\,K. In all systems, the superfluid stiffness $\rho_s(T)=\lambda^2(0)/\lambda^2(T)$ can be approximated by a power-law $T$ dependence at low temperatures, indicating the robustness of strong momentum anisotropy in the superconducting gap $\Delta(\bm{k})$. The power $\alpha$ increases from $\sim1$ with mass enhancement and approaches an unconventional exponent $\alpha\sim 1.5$ in the heaviest CsFe$_2$As$_2$. This appears to be a hallmark of superconductors near antiferromagnetic quantum critical points, where the quasiparticles excited across the anisotropic $\Delta(\bm{k})$ are significantly influenced by the momentum dependence of quantum critical fluctuations. 

\end{abstract}



\maketitle


An important question on the mechanism of high-temperature superconductivity in iron pnictides concerns the relationship between the quantum criticality and superconductivity. In the vicinity of the range of the superconducting phase, antiferromagnetic order with a close-by tetragonal-orthorhombic structural transition is frequently found, and the anomalous normal-state properties near the endpoint of the order have been detected in BaFe$_2$As$_2$-based superconductors, implying the importance of quantum critical fluctuations \cite{Shibauchi14,Kasahara10,Walmsley13,Nakai13,Yoshizawa12,Chu12,Gallais13}. These anomalies are mostly associated with the enhanced electron correlations, which manifest themselves as divergently increasing electron effective mass $m^*$ approaching a quantum critical point (QCP) \cite{Walmsley13}. 

\begin{figure}[tbp]
\includegraphics[width=1\linewidth]{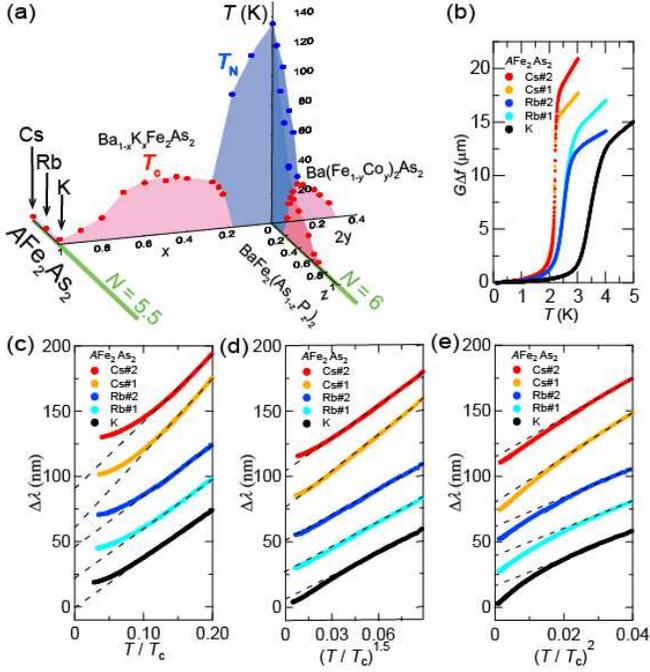}
\vspace{-3mm}
\caption{(Color online). (a) Schematic phase diagrams of BaFe$_2$As$_2$-based superconductors with Co, K, and P substitutions \cite{Shibauchi14}, combined with that of  $A$Fe$_2$As$_2$ ($A=$ K, Rb, and Cs). Only antiferromagnetic $T_N$ (blue circles) and superconducting transition temperatures $T_c$ (red circles) are shown for clarity. The isovalent  $A$Fe$_2$As$_2$ and BaFe$_2$(As$_{1-z}$P$_z$)$_2$ systems correspond to $3d$ electron numbers per Fe atom of $N=5.5$ and 6.0, respectively (green lines). (b) Temperature dependence of the frequency shift multiplied by the geometric factor $G$ in $A$Fe$_2$As$_2$ single crystals. The data for $A=$ K are taken from Ref.\,\cite{Hashimoto10}. (c)-(e) Low-temperature change of the penetration depth $\Delta\lambda$ below $T/T_c=0.20$ plotted against $T/T_c$ (c), $(T/T_c)^{1.5}$ (d), and $(T/T_c)^{2}$ (e). The data are vertically shifted for clarity. Dashed lines are the guides to the eyes.
}
\label{lambda}
\end{figure}

On the other hand, in the high doping regime of the Ba$_{1-x}$K$_x$Fe$_2$As$_2$ system, an increasing trend of $m^*$ has been found with hole doping $x$; the Sommerfeld coefficient of the electronic specific heat $\gamma$, a measure of $m^*$, reaches $\sim 100$\,mJ/mol\,K$^2$ in KFe$_2$As$_2$ ($x=1$) \cite{Fukazawa11,Hardy13,Storey13}, and even exceeds $\gamma$ of the quantum critical concentration $z=0.3$ in BaFe$_2$(As$_{1-z}$P$_z$)$_2$ system \cite{Walmsley13}. Such a very large mass in KFe$_2$As$_2$, comparable to moderately heavy fermion materials, is also confirmed by the large Fermi-liquid coefficient of the resistivity \cite{Hashimoto10}, and by quantum oscillations \cite{Terashima13},  angle-resolved photoemission \cite{Yoshida14}, and optical conductivity \cite{Nakajima14}. It has been theoretically suggested that these strongly enhanced electron correlations for the $x=1$ system, where the number $N$ of $3d$ electrons per Fe site is 5.5, are related to the possible proximity to the Mottness of $N=5$ (half-filled for all $3d$ orbitals) \cite{Medici14}, which is much more pronounced than the $N=6$ case of BaFe$_2$As$_2$ \cite{Gerorges13,Yu13}. It has also been found that the isovalent substitution for K by larger alkali-ions Rb or Cs results in even larger  $\gamma$ values up to  $\sim 180$\,mJ/mol\,K$^2$ \cite{Wang13,Zhang15}. This suggests that the negative chemical pressure effect in $A$Fe$_2$As$_2$ ($A=$ K, Rb, and Cs) [see Fig.\:\ref{lambda}(a)] effectively reduces the band width with increasing alkali-ion radius, which leads to an approach to a QCP of another antiferromagnetically ordered phase \cite{Zocco15}. 

One of the key issues is how the quantum critical fluctuations affect the superconducting properties 
\cite{Hashimoto12,Hashimoto13,Chowdhury13,Levchenko13,Nomoto13,Putzke14,Lamhot15}.  It has been demonstrated both experimentally \cite{Hashimoto12,Lamhot15} and theoretically \cite{Chowdhury13,Levchenko13,Nomoto13} that the enhanced electron correlations lead to an enhancement of the London penetration depth $\lambda(0)$ in the superconducting phase. It has also been proposed \cite{Hashimoto13,Nomoto13} that the low-energy quasiparticle excitations in nodal superconductors near an antiferromagnetic QCP are significantly affected in a non-trivial way, giving rise to an unusual power-law temperature dependence of $\lambda(T)$ with a non-integer exponent close to 1.5, which differs distinctly from the $T$-linear dependence expected for line nodes of the superconducting energy gap $\Delta(\bm{k})$. In this viewpoint, clarifying the effect of the possible QCP on $\lambda(T)$ in $A$Fe$_2$As$_2$ ($A=$ K, Rb, and Cs) is fundamentally important. Here, we report on such measurements of $\lambda(T)$ down to low temperatures by using very clean single crystals. We find a systematic change in the exponent $\alpha$ of its power-law temperature dependence with increasing alkali-ion radius, from $\alpha\sim 1$ for K \cite{Hashimoto10} to $\alpha\sim 1.5$ for Cs. This supports the presence of quantum critical fluctuations in CsFe$_2$As$_2$, strongly affecting quasiparticle excitations as observed in other quantum critical superconductors.


Single crystals of $A$Fe$_2$As$_2$ were synthesized in alumina crucibles by a self-flux method \cite{Zocco13}. In these crystals, quantum oscillations have been observed in magnetostriction measurements for $A=$ K \cite{Zocco14}, Rb, and Cs \cite{Zocco15}, indicating that the samples are very clean. In this study, the penetration depth measurements were performed for $A=$ Rb, and Cs using a self-resonant tunnel-diode oscillator with the resonant frequency $f$ of $\sim 13$\,MHz in a dilution refrigerator, and we compare the results with the previously reported data for $A=$ K taken in a similar setup \cite{Hashimoto10}. The shift of the resonant frequency $\Delta f$ is directly proportional to the change in the magnetic penetration depth, $\Delta\lambda(T)$ = $G\Delta f(T)$, in the superconducting state below $T_c$. The geometric factor $G$ is determined from the geometry of samples and the coil. The samples are cleaved on all sides to avoid degradation due to the reaction with air and coated by Apiezon grease. The samples were mounted in the cryostat within 15 minutes after cleavage to minimize the exposure to air. Typical lateral size of crystals is $\sim 500 \times 500$\,$\mu$m$^2$ and the thickness is less than 50\,$\mu$m.

Figure\:\ref{lambda}(b) shows the temperature dependence of the resonant frequency shift $\Delta f$ multiplied by $G$. The superconducting transition temperatures $T_c$ defined by the midpoint of the transition are 3.4, 2.5, and 2.2\,K for $A=$ K, Rb, and Cs, respectively, consistent with the previous studies \cite{Shermadini10, Hong13, Wang13, Tafti15, Wu15}. In the normal state above $T_c$, $G\Delta f$ is determined by the skin depth $\delta(T)$, which is related to the dc resistivity $\rho(T)$ through $\delta(T) = \sqrt{2\rho(T)/\mu\omega}$. Here,  $\mu$ is the permeability and $\omega=2\pi f$ is the angular frequency of the oscillator. From these relations we estimate $\rho$ values just above $T_c$ as $\sim 0.8$, $\sim 1.0$, and $\sim 1.2$\,$\mu\Omega$\,cm for the K, Rb, and Cs cases, respectively. These low resistivity values indicate the high quality of our samples.

In Figs.\:\ref{lambda}(c)-(e), the change in the penetration depth $\Delta\lambda(T) = \lambda(T)-\lambda(0)$ at low temperatures is plotted against $T/T_c$, $(T/T_c)^{1.5}$ and $(T/T_c)^2$ for all samples. Two crystals are measured for both Rb and Cs systems and they exhibit almost identical $T$-dependence in each case. For all systems $\Delta\lambda(T)$ clearly exhibits a non-exponential $T$-dependence in a wide temperature range, which can be approximated by a power law. The exponent of the power-law $T$-dependence is less than 2, which is clearly demonstrated by the convex curvature against $(T/T_c)^2$ [Fig.\:\ref{lambda}(e)]. Such small exponents $\alpha<2$ found for all three compounds, which can hardly be explained by the pair-breaking scattering effect for fully opened gaps \cite{Prozorov11}, immediately indicates the robust presence of low-energy quasiparticle excitations. This is consistent with nodal gap structure inferred from the observation of a residual term $\kappa/T$ for $T\to 0$\,K of the thermal conductivity $\kappa$ in $A$Fe$_2$As$_2$ \cite{Dong10,Reid12,Watanabe14,Zhang15,Hong13}, although measurements at lower temperature below $\sim0.1$\,K would be required for the ultimate determination of the presence of nodes or very small minima in $\Delta(\bm{k})$ \cite{Hardy14,Carrington15}. 

\begin{figure}[tbp]
\includegraphics[width=1.0\linewidth]{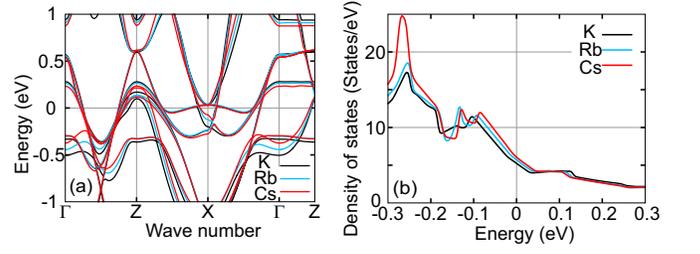}
\vspace{-3mm}
\caption{(Color online). (a) Band structures of $A$Fe$_2$As$_2$ ($A=$ K, Rb, and Cs) calculated within GGA-PBE. (b) Density of states as a function of energy near the Fermi level.
}
\label{band}
\end{figure}

A closer look at the data shows a systematic increase of the exponent in the order of K, Rb, and Cs. To discuss the origin of this change, we first examine the evolution of electronic structure in this series. The band-structure calculations are performed within the generalized gradient approximation (GGA-PBE \cite{Perdew96})  for the experimentally obtained lattice parameters by using the Wien2k package \cite{Blaha}. As shown in Fig.\:\ref{band}(a), the results indicate very similar structures for all three systems, which are consistent with the local density approximation results for $A=$ K \cite{Terashima13} and Cs \cite{Kong14}. The Fermi surfaces of the K, Rb, and Cs systems all have the same topology, which is confirmed by the quantum oscillation experiments mentioned above \cite{Zocco15}. Furthermore, $T_c$ of these materials shows a universal V-shaped dependence on hydrostatic pressure \cite{Tafti15}, suggesting that the superconducting gap structure is also essentially similar.

\begin{figure}[tbp]
\includegraphics[width=1\linewidth]{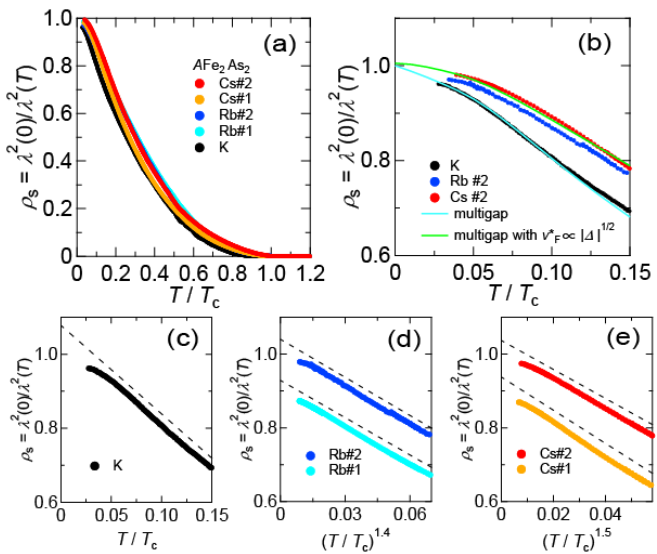}
\vspace{-3mm}
\caption{(Color online). (a) Superfluid stiffness $\rho_s(T) = \lambda^2(0)/\lambda^2(T)^2$ obtained by using the reported values of $\lambda(0)=203$ \cite{Kawano11}, 267\,nm \cite{Shermadini10}, and the estimate of $\lambda(0)=305$\,nm (see text) for $A=$ K, Rb, and Cs, respectively.  (b)  Comparisons of $\rho_s(T)$ for $T/T_c<0.2$ with multigap calculations with and without considering the momentum-dependent effective Fermi velocity $v_F^*\propto|\Delta(\bm{k})|^{1/2}$. (c)-(e) Low-$T$ part of $\rho_s(T)$ below $T/T_c=0.15$ plotted against $T/T_c$ for K (c), $(T/T_c)^{1.4}$ for Rb (d), and $(T/T_c)^{1.5}$ for Cs (e). The data are vertically shifted for clarity. Dashed lines are the guides for the eyes.
}
\label{rho_s}
\end{figure}

To make a more quantitative analysis, we use the value of $\lambda(0) \sim 203(10)$\,nm reported in the small-angle neutron scattering measurements for K \cite{Kawano11} and $\lambda(0) \sim 267(5)$\,nm obtained in the $\mu$SR measurement for Rb \cite{Shermadini10}. As there is no report on $\lambda(0)$ for Cs so far, we roughly estimate $\lambda(0) \sim 305(30)$\,nm on the assumption that $\lambda^2(0)$ is proportional to $\gamma$ \cite{Walmsley13}. By using these values, we obtain the full temperature dependence of $\lambda(T)$ and the normalized superfluid stiffness $\rho_s(T) = \lambda^2(0)/\lambda^2(T)$ [Fig.\:\ref{rho_s}(a)]. The overall $T$ dependence of $\rho_s$ is quite similar in all samples, consistent with the very similar electronic and gap structures in these materials except for the effective mass, which is mainly reflected by the difference of the $\lambda(0)$ values. 
In contrast, $\rho_s(T)$ at low temperatures shows a drastic change; compared with the data for K, the Rb and Cs data show more convex curvatures [Fig.\:\ref{rho_s}(b)], indicating a significant change in the amount of quasiparticle excitations at low energies. 

The low-temperature part of $\rho_s(T)$ plotted against $(T/T_c)^\alpha$ with different exponents $\alpha$ for different systems [Figs.\:\ref{rho_s}(c)-(e)] confirms a power-law $T$-dependence with a systematic change in $\alpha$. In particular, we find $1-\rho_s(T)\propto T^{\alpha}$ with $\alpha \sim 1.5$ for Cs in a wide $T$ range, which is clearly different from $\alpha \sim 1$ for K. It should be noted that there is a small deviation from the power-law dependence at the lowest temperatures below $\sim 0.15$\,K, which may be due to impurity scattering. Such disorder effects can induce $T^2$ dependence of penetration depth at sufficiently low temperatures \cite{Hirschfeld93,Bonn94}, and sometimes open a small gap in $s$-wave superconductors with accidental nodes \cite{Mizukami14}, but these effects are limited to the correspondingly low temperature scale and will not affect the higher-$T$ behavior, which is determined by the intrinsic spectrum of quasiparticle excitations. To describe the disorder effect in superconductors with line nodes, the empirical formula $\Delta\lambda(T) \propto (T/T_c)^2/\{(T/T_c)+(T^*\!/T_c)\}$ that interpolates $T$-linear and $T^2$ dependences has been widely used, where $T^*$ is a measure of the impurity scattering rate \cite{Hirschfeld93}. However, this fitting procedure to our data in a wide temperature range up to $T/T_c=0.2$ gives considerably large values of $T^*\!/T_c$; the power-law dependence with $\alpha\sim1.5$ in CsFe$_{2}$As$_{2}$ can be approximated by the above equation with $T^*\!/T_c\sim 0.9$ which is unphysically large for clean samples with low residual resistivity \cite{Hashimoto13}. Such a large  $T^*\!/T_c$ is also incompatible with the nonlocality effect, in which $T^*\!/T_c \sim \xi(0)/\lambda(0) \ll 1$ where $\xi$ is the coherence length \cite{Kosztin97}. Therefore we conclude that the anomalous power $\alpha\sim 1.5$ observed for the Cs system is an intrinsic property of the low-energy quasiparticles excited near the nodes or minima of $\Delta(\bm{k})$.  

\begin{figure}[tbp]
\includegraphics[width=0.7\linewidth]{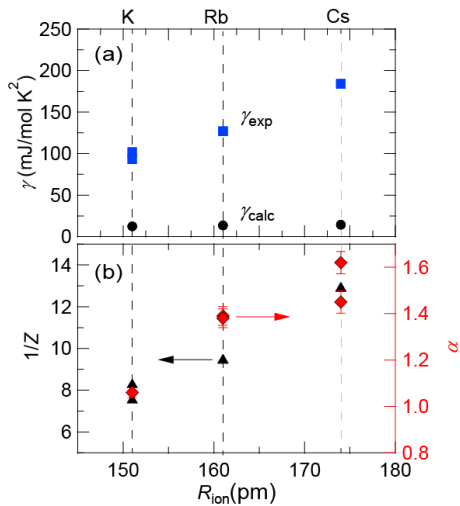}
\vspace{-3mm}
\caption{(Color online). (a) Sommerfeld coefficients $\gamma_{\rm exp}$ in single crystals of $A$Fe$_2$As$_2$ ($A=$ K, Rb, and Cs) \cite{Fukazawa11,Hardy13,Wang13,Zhang15}, compared with the estimated $\gamma_{\rm calc}$ from the band-structure calculations, as a function of alkali-ion radius $R_{\rm ion}$ in the eight-fold coordination. (b) Mass renormalization factor $1/Z$ extracted from the comparisons of experimental and calculated $\gamma$ values, and the exponent $\alpha$ of the power-law $T$ dependence of $\rho_s(T)$ in a temperature range up to $T/T_c=0.15$.
}
\label{trends}
\end{figure}

The most pronounced change with the alkali-ion radius $R_{\rm ion}$ in this series is the strong enhancement of the experimentally determined Sommerfeld coefficient $\gamma_{\rm exp}$, as shown in Fig.\:\ref{trends}(a). This $R_{\rm ion}$ dependence of $\gamma_{\rm exp}$ is much more rapid than that of the band-structure calculations $\gamma_{\rm calc}$, which changes by only $\sim 15$\% as can be seen in the small increase in the density of states at the Fermi level [Fig.\:\ref{band}(b)]. This indicates that the mass renormalization factor averaged over the Fermi surface, $1/Z=\gamma_{\rm exp}/\gamma_{\rm calc}$, is strongly enhanced with increasing $R_{\rm ion}$ [Fig.\:\ref{trends}(b)], which is consistent with the analysis of quantum oscillations \cite{Zocco15}. The value of $1/Z$ in CsFe$_2$As$_2$ reaches $\sim13$, which is larger than the estimated value of $\sim10$ for the quantum critical concentration $z=0.3$ in the BaFe$_2$(As$_{1-z}$P$_z$)$_2$ system \cite{Walmsley13}. This result suggests that negative chemical pressure brings the $A$Fe$_2$As$_2$ system toward a QCP. The exponent $\alpha$  obtained from the power-law fit to the $1-\rho_s(T)$ data in the $T$ range up to $0.15 T_c$ also shows a systematic change with $R_{\rm ion}$ as shown in Fig.\:\ref{trends}(b).  This highlights a close link between the unconventional exponent $\sim1.5$ and quantum criticality. Considering the fact that $\alpha\sim1.5$ has been reported for several unconventional superconductors in the vicinity of QCPs \cite{Hashimoto13}, we infer that this anomaly might have a common origin associated with quantum critical fluctuations.

It has been proposed \cite{Hashimoto13} that an unusual power-law dependence of $\rho_s(T)$ with a non-integer exponent might arise from a strong momentum dependence of the effective Fermi velocity $v_F^*(\bm{k})$ near the nodes of the superconducting gap $\Delta(\bm{k})$. Such a $\bm{k}$ dependence of $v_F^*\propto 1/m^*$ requires that quantum critical fluctuations responsible for the mass renormalization also have strong momentum dependence associated with $\Delta(\bm{k})$. This can be naturally expected when the quantum critical fluctuations are quenched by opening the superconducting gap, the degree of which is determined by the gap magnitude $|\Delta(\bm{k})|$ \cite{Hashimoto13}. Recent theoretical calculations for quantum critical superconductors with line nodes \cite{Nomoto13} indicate that the current vertex corrections to the $T$-linear penetration depth can be significant if the antiferromagnetic hot spots are located near the nodal points such as is the case of electron-doped cuprates. 

Now we extend this consideration to the multigap systems, where $\Delta(\bm{k})$ is very different for different bands. The strong curvature found in $\rho_s(T)$ near $T_c$ indicates the importance of the multiband effect \cite{Prozorov11}. Indeed recent ARPES and specific-heat measurements have reported highly band-dependent gap structures in KFe$_{2}$As$_{2}$ \cite{Okazaki12,Hardy14}. As shown in  Fig.\:\ref{rho_s}(b), we find that a set of multigap functions similar to that accounting for the specific-heat data \cite{Hardy14} can reproduce the low-temperature $\rho_s(T)$ of KFe$_2$As$_2$ fairly well. Here we calculate  $\rho_s(T)=\sum_i{w^i\rho_s^i}(T)$ by assuming an extended $s$-wave state with a nodal gap in one band (the middle hole band $\zeta$ around $\Gamma$) and different constant gaps in other bands. The weight $w^i$ of each band $i$ is estimated from the quantum oscillation results \cite{Terashima13} (see Table\:\ref{multigap}). Apparently, these weights do not change significantly for Rb and Cs \cite{Zocco15}, and thus we need another ingredient to explain the enhanced exponent $\alpha$. To include the effect of quantum criticality discussed above, we then calculate $\rho_s(T)$ by applying the $\bm{k}$-dependent effective velocity in the superconducting state of the form $v_F^*(\bm{k})\propto|\Delta(\bm{k})|^{1/2}$ \cite{Hashimoto13} to all bands, which affects the weight that includes the effective mass \cite{SI}. For the $\zeta$ band this $\bm{k}$-dependence increases the low-$T$ power-law exponent of $\rho_s^{\zeta}(T)$. For the other bands, the $T$ dependence of each $\rho_s^i$ is unchanged but the relative weights are modified by their gap magnitudes; for the small-gap band the weight is reduced because of the suppressed $v_F^*$. This simple analysis captures salient features of the systematic change in $\rho_s(T)$ with $A$ [Fig.\:\ref{rho_s}(b)], although at higher temperatures there are some deviations which are possibly due to the interband coupling effect. The essence of this $|\Delta(\bm{k})|$-dependent mass renormalization is that it can naturally reduce the contribution of quasiparticles excited at the positions where the gap is small. Thus the observed positive correlation between the mass enhancement and the exponent of the superfluid stiffness supports this non-trivial effect of quantum critical fluctuations in the superconducting state.

%


\begin{table}[tbp]
 \begin{center}
  \caption{Multigap parameters used in the calculations of $\rho_s(T)$. The weight $w^i$ to $\rho_s$ of band $i$ is estimated from the number of holes $n$ and the effective mass $m^*$ in units of the electron mass $m_e$ \cite{Hashimoto10}, which are determined by the quantum oscillations in KFe$_2$As$_2$ \cite{Terashima13}. The gap values of each band and the angle dependence of $\Delta(\phi)$ of $\zeta$ band are assumed following Ref.\,\cite{Hardy14}.}
   \begin{tabular}{cccccc} 
\hline 
\hline
band & $n$ & ${m^*}/{m_e}$ & $w^i$ & $\Delta(\phi)/k_BT_c$ \\
\hline
$\alpha$ (inner) & 0.17 & 6 & 0.31 & 0.61 \\ 
$\zeta$ (middle) & 0.26 & 13 & 0.23 &  $0.35+0.45\cos(4\phi)$ \\ 
$\beta$ (outer) & 0.48 & 18 & 0.31 & 0.25 \\ 
$\varepsilon$ (corner) & 0.09 & 7 & 0.15 & 1.8 \\ 
total & 1.0 &  & 1.0 &  \\ 
\hline
\hline
   \end{tabular}
   \label{multigap}
 \end{center}
\end{table}


In summary,  from the penetration-depth measurements of $A$Fe$_2$As$_2$ ($A=$ K, Rb, and Cs) in the heavily hole-doped regime with $3d$ electron number of $N=5.5$, we show that the superconducting gaps have robust strong anisotropies, consistent with the previous reports of thermal conductivity measurements  \cite{Dong10,Reid12,Watanabe14,Zhang15,Hong13}. The exponent of the approximate power-law dependence of the superconducting stiffness changes from $\alpha\sim 1$ toward $\alpha\sim 1.5$ with negative chemical pressure, indicating a systematic decrease of the low-energy quaiparticle excitations which correlates with the enhancement of the effective mass toward the possible QCP in $A$Fe$_2$As$_2$. This observation entails a new relation between the momentum dependencies of quantum critical antiferromagnetic fluctuations and quasiparticle excitations of the superconducting condensate. 


We thank A. Carrington, F. Eilers, F. Hardy, R. Heid, P.\,J. Hirschfeld, and C. Meingast for valuable discussions. This work has been supported by 
Bilateral Joint Research Project program and KAKENHI from JSPS.



\widetext

\newpage


\renewcommand{\tablename}{TABLE S$\!$}
\renewcommand{\figurename}{FIG. S$\!\!$}
\renewcommand{\theequation}{S\arabic{equation}}
\setcounter{table}{0}
\setcounter{figure}{0}
\setcounter{equation}{0}

\section{Supplemental Material}

\subsection{Multigap calculations of the superfluid stiffness}

The London penetration depth in a superconductor can be calculated from the superconducting gap $\Delta(\bm{k},T)$ through the following equation\cite{Hashimoto13}:
\begin{equation}\label{Eq:lam}
\lambda_{jj'}^{-2}(T)=
\frac{\mu_0e^2}{4\pi^3\hbar}\int \frac{v_{Fj}(\bm{k})v^*_{Fj'}(\bm{k})}{|\bm{v}_{F}(\bm{k})|}
\left[1-Y(\bm{k},T)\right]
d\bm{S},
\end{equation}
where $v^*_F$ ($v_F$) is the effective velocity in the superconducting (normal) state, the subscripts $j,j'$ denote the directions of the current and vector potential. 
Here, 
\begin{equation}
Y(\bm{k},T)=
-2 \int_{\Delta(\bm{k},T)}^\infty \frac{\partial f(E)}{\partial E}\frac{E}{\sqrt{E^2-\Delta^2(\bm{k},T)}} dE
\end{equation}
is the Yosida function and $f(E)$ is the Fermi-Dirac function for the quasiparticle energy $E$. In this study, we measure the in-plane penetration depth $\lambda(T)$ and thus consider the in-plane effective Fermi velocities. 

In multiband superconductors, the superfluid stiffness can be calculated by adding the contributions from different bands as
\begin{equation}
\rho_s(T)=\frac{\lambda^{2}(0)}{\lambda^{2}(T)}=\sum_i{w^i\rho_s^i(T)}, 
\end{equation}
where band $i$ has the superfluid stiffness $\rho_s^i(T)$. The weight $w^i$ to the total $\rho_s(T)$ is determined by the corresponding carrier concentration divided by its effective mass.
The weight of each band can be estimated from the analysis of quantum oscillations, which yields information on the cross-sectional area and the effective mass $m^*$ of the observed extremal orbits. Importantly the magnitude of the effective mass averaged within one band affects only the absolute value of $\lambda$ as well as the weight $w^i$ but does not affect the temperature dependence of $\lambda$ as one can see in Eq.\:(S1). 

Based on the quantum oscillation results \cite{Terashima13}, we estimate the weight $w^i$ as listed in Table\:I for KFe$_2$As$_2$ \cite{Hashimoto10}. Here we ignore the $\delta$ band near the $Z$ point, whose contribution is very small. The calculated contributions $w^i\rho_s^i(T)$ and the total $\rho_s(T)$ are shown in Fig.\:S1(a), together with the experimental data of KFe$_2$As$_2$ \cite{Hashimoto10}. Here the gap function $\Delta(\bm{k})$ of each band is assumed as suggested by Hardy {\it et al.}  \cite{Hardy14} (see Table\:I). Only the $\zeta$ band does have a wave form with nodes at eight directions as found in high-resolution laser ARPES measurements \cite{Okazaki12}. For the other bands we assume constant $\Delta$ values with large differences in magnitude. As we are interested in $\rho_s(T)$ at low temperatures, we simply assume the BCS-type temperature dependence of the gap which vanishes at the same $T_c$ for all bands. This assumption will introduce some errors at high temperatures when the interband coupling is strong \cite{Prozorov11,Hardy14}, but the low-temperature behavior should not be affected by the inteband effect.  As expected, $\rho_s^i(T)$ of the $\varepsilon$ band, whose gap magnitude is largest, is very flat at low temperatures, and $\rho_s^i(T)$ of the $\zeta$ band with a nodal gap has the steepest temperature dependence. Interestingly, $\rho_s^i(T)$ of the $\beta$ band with a very small gap also shows very strong temperature dependence at low $T$. The total $\rho_s(T)$  obtained by adding $\rho_s^i(T)$ with different $T$ dependences essentially reproduces the salient features of the experimental data for KFe$_2$As$_2$, indicating that the set of gap parameters suggested in \cite{Hardy14} can account for both the sepecific heat and low-temperature penetration depth in a semi-quantitative manner. 

Having established the validity of the multigap fit for KFe$_2$As$_2$, we now consider the power-law exponent found in CsFe$_2$As$_2$. The quantum oscillation experiments for the $A$Fe$_2$As$_2$ series \cite{Zocco15} indicate that the size of the Fermi surface of each band is almost unchanged and that the effective mass of each band is nearly equally enhanced with alkali-ion radius.
These features suggest that $w^i$ does not change much. To model the effect of the quantum critical mass enhancement in the superconducting state, we take the same set of superconducting gaps, yet with the momentum dependence of $v^*_F(\bm{k})\propto |\Delta(\bm{k})|^{1/2}$, which has been suggested to account for the unconventional exponent of $\alpha\sim1.5$ in nodal quantum critical superconductors by Hashimoto {\it et al.} \cite{Hashimoto13}. When we extend this model to multigap systems, we expect that it will also affect the relative weight, i.e., ; $w^i/w^{i'}$ can be replaced by $|\Delta^i|^{1/2}w^i/|\Delta^{i'}|^{1/2}w^{i'}$ (where $|\Delta^i|$ is the gap magnitude of band $i$). For the $\zeta$ band, $\rho_s^i(T)$ also changes because the $\bm{k}$ dependence of $v_F^*$ affects the integration in Eq.\:(S1). 
The resulting $\rho_s^i(T)$ calculations for each band and the total $\rho_s(T)$ for CsFe$_2$As$_2$ are show in Fig.\:S1(b) together with the experimental data. The difference of $\rho_s^{\zeta}(T)$ for the nodal gap $\Delta(\phi)/k_BT_c=0.35+0.45\cos(4\phi)$ between constant $v^*_F$ and  $v^*_F(\bm{k})\propto |\Delta(\bm{k})|^{1/2}$ can be clearly seen in Fig.\:S1(c). This momentum-dependent mass renormalization is directly affected by the superconducting gap magnitude, which effectively reduces the $\rho_s$ contribution of quasiparticles excited at $\bm{k}$ vectors where the gap is small. This mechanism can account for the systematic changes in the low-energy quasiparticle excitations when the system approaches a quantum critical point in the vicinity of a superconducting phase.

\begin{figure}[htbp]
\includegraphics[width=0.8\linewidth]{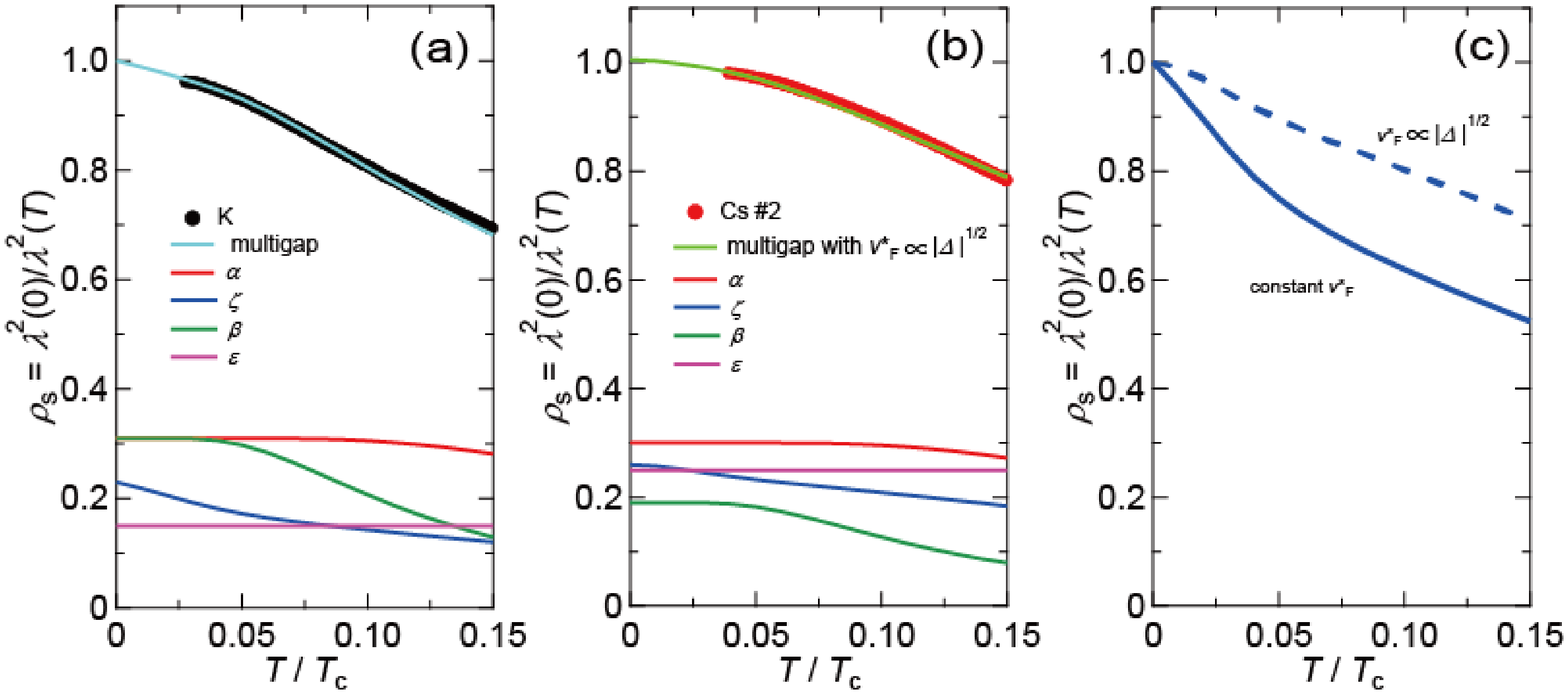}
\vspace{-3mm}
\caption{(a) Low-temperature superfluid stiffness of KFe$_2$As$_2$ (circles) with the multigap fit. Contributions from different bands calculated by the parameters given in Table\:I are also shown (solid lines).  (b) Low-temperature superfluid stiffness of CsFe$_2$As$_2$ (circles) with the multigap fit considering the effect of quantum criticality assuming the momentum dependence of  $v^*_F(\bm{k})\propto |\Delta(\bm{k})|^{1/2}$. The zero-temperature values of each curve correspond to the effective weight taking into account the difference of gap magnitude. (c) Calculated results of $\rho_s^{\zeta}(T)$ with and without considering $\bm{k}$-dependence of $v^*_F$ in the superconducting state. 
}
\label{multigap}
\end{figure}

%



\end{document}